\documentclass[12pt]{article}

\newcommand{\be}{\begin{equation}}
\newcommand{\ee}{\end{equation}}
\newcommand{\bea}{\begin{eqnarray}}
\newcommand{\eea}{\end{eqnarray}}
\newcommand{\nb}{\nolinebreak}
\newcommand{\gs}{\ensuremath{g_s}}      
\newcommand{\ap}{\ensuremath{\alpha'}}  
\newcommand{\ls}{\ensuremath{l_s}}      
\newcommand{\lP}{\ensuremath{l_P}}      
\newcommand{\Fp}{\ensuremath{F'}}
\newcommand{\Fe}{\ensuremath{F_{\epsilon}}}
\newcommand{\eps}{\ensuremath{\epsilon}}
\newcommand{\xp}{\ensuremath{x'^{+}}}
\newcommand{\xm}{\ensuremath{x'^{-}}}

\newcommand{\Pm}{\ensuremath{P'_{-}}}
\newcommand{\Pup}{\ensuremath{{P'}^{+}}}
\newcommand{\Pum}{\ensuremath{{P'}^{-}}}
\newcommand{\calA}{\ensuremath{\mathcal{A}}}
\newcommand{\calB}{\ensuremath{\mathcal{B}}}
\newcommand{\calI}{\ensuremath{\mathcal{I}}}
\newcommand{\calO}{\ensuremath{\mathcal{O}}}
\newcommand{\calS}{\ensuremath{\mathcal{S}}}
\newcommand{\tgs}{\ensuremath{\tilde{\gs}}}
\newcommand{\tls}{\ensuremath{\tilde{\ls}}}
\newcommand{\tlP}{\ensuremath{\tilde {\lP}}}
\newcommand{\tR}{\ensuremath{\tilde{R}}}

\input epsf                    

\begin{document}

\begin{titlepage}

\title{
\begin{flushright}
\begin{small}
hep-th/9804034\\
PUPT-1784  \\
\end{small}
\end{flushright}
\vspace{1.cm}
Is Physics in the Infinite Momentum
Frame Independent of the Compactification Radius?}
\author{Alberto G\"uijosa\\
{\small Joseph Henry Laboratories}\\ 
{\small Princeton University}\\
{\small Princeton, NJ 08544 USA}\\
{\tt \small aguijosa@princeton.edu}}
\date{}
\maketitle

\begin{abstract} 
 
 With the aim of clarifying the eleven dimensional content of Matrix 
 theory, we examine the dependence of a theory in the infinite momentum
 frame (IMF) on the (purely spatial) longitudinal compactification 
 radius $R$.
 
 It is shown that in a point particle theory the generic scattering
 amplitude becomes {\em independent} of $R$ in the IMF. Processes with zero 
 longitudinal momentum transfer are found to be exceptional. 
 The same question is addressed in a theory with extended objects.
 A one-loop type II string amplitude is shown to be $R$-independent 
 in the IMF, and to coincide with that of the uncompactified theory.  
 No exceptional processes exist in this case.

 The possible implications of these results for M-theory are discussed. In
 particular, if amplitudes in M-theory are independent of $R$ in the
 IMF, Matrix theory can be rightfully expected (in the $N\to\infty$ limit) 
 to describe {\em uncompactified}\/ M-theory.

\end{abstract}
\vspace{1cm}
\centerline{PACS codes: 11.25.w, 11.25.Db, 11.80.m, 11.30.Cp}
\vspace{2cm}
\hspace*{1.1cm}June 1998

\thispagestyle{empty}
\end{titlepage}

\section{Introduction}

 Dramatic advances in recent years have led to the realization that 
 a consistent eleven dimensional quantum theory
 of gravity, M-theory, should exist \cite{M}.
 Over a year ago Banks, Fischler, Shenker and Susskind \cite{BFSS}
 (BFSS) made the bold conjecture that M-theory in the infinite 
 momentum frame (IMF) \cite{Wein} has a precise description as a particular
 limit of a matrix quantum mechanical system. The system in
 question was originally obtained 
 as a description of D0-brane physics in {\em ten dimensional} string 
 theory \cite{Wi,DKPS}. 
 
 For the definition of the
 model, the longitudinal direction is
 compactified on an auxiliary (spatial) circle of radius $R$; the total
 longitudinal momentum is then quantized, $P=N/R$.
 The physics of uncompactified M-theory is expected to be recovered
 in the $N,R,N/R\to\infty$ limit. Several
 rather remarkable pieces of evidence were presented for the 
 conjecture in \cite{BFSS}, and others 
 followed\footnote{See \cite{Ba} and \cite{Sulect} for reviews of what has 
 been accomplished by the Matrix model, along with extensive references.},
 but progress was hampered by the 
 technical difficulty of studying the large $N$ limit.
 
 The situation improved last year when
 Susskind \cite{Su} conjectured that even the finite $N$ Matrix model
 should have direct physical meaning as the discrete light cone 
 quantization (DLCQ) \cite{DLCQ} of M-theory.  
 In that case the description is in terms of an ordinary 
 (as opposed to infinite momentum) reference frame, and the compact 
 direction is null, with radius $R'$. This form of the conjecture allowed  
 several more stringent tests, by comparison of the Matrix amplitudes
 with those in the DLCQ of supergravity. If Susskind's conjecture 
 were correct, the $N,R'\to\infty$ limit of the model, 
 with $N/R'$ fixed,
 would be expected to yield the standard 
 (uncompactified) light front quantization (LFQ) of M 
 theory\footnote{Since in the case of a lightlike
 compactification $R'$ can be rescaled by longitudinal boosts, one 
 would also expect to recover uncompactified physics in the Lorentz
 equivalent limit $N\to\infty$, with $R'$ fixed.}.
 
 In another important paper, Seiberg \cite{Se} constructed a 
 seemingly miraculous proof that 
 the finite $N$ Matrix model indeed describes the DLCQ of M-theory.
 This, however, was done at the cost of defining DLCQ as being Lorentz
 related to an IMF compactification on a {\em vanishingly small}
 spatial circle.
 As a result, the question of whether the model captures 
 eleven dimensional physics is greatly obscured.
 The large $N$ limit of such a DLCQ would be equivalent to the BFSS 
 limit, but with a spatial radius $R \to 0$ instead of $R \to \infty$.
 We will henceforth refer to Seiberg's limit as `near DLCQ' to 
 distinguish it from the conventional DLCQ \cite{DLCQ}. That this is the 
 useful way to interpret what is meant by DLCQ in the context of the 
 Matrix model is evidenced by the success of \cite{Se} (see 
 also \cite{Sen}) in providing a uniform prescription for toroidal 
 compactifications.
 
 At about the same time, some discrepancies between Matrix model and
 supergravity scattering amplitudes were found \cite{DR}. 
 (Difficulties for Matrix model amplitudes in nontrivial 
 backgrounds have also been reported \cite{DOS}.)
 The status of the model is therefore at present uncertain, 
 and there has been some controversy in the literature 
 regarding what it is exactly that Seiberg proved, and whether or not
 Matrix theory and supergravity amplitudes should be directly compared.
 
 To gain additional insight, Hellerman and Polchinski \cite{HP} examined the 
 limit of a field theory compactification on an almost lightlike circle (or 
 equivalently, on a vanishingly small spatial circle in the IMF),
 at fixed $N$. They found the limit to be complicated:
 the longitudinal zero modes become strongly coupled, and as a result,
 in almost all theories some of the perturbation theory amplitudes diverge.
 As a consequence of this, the authors of \cite{HP} expressed 
 serious doubts regarding the relevance
 of the Matrix model for describing uncompactified M-theory. In 
 particular, they emphasized that amplitudes in finite $N$ Matrix theory
 and supergravity do not have, {\it a priori,}\/
 a common range of validity.
 
 De Alwis \cite{deAl} has used the scaling limits 
 of \cite{Se} to argue that finite $N$ Matrix model and supergravity 
 amplitudes (with one impact parameter and no longitudinal
 momentum transfer) should be expected to agree as 
 a result of string world-sheet duality.
 
 More recently, Kabat and Taylor \cite{KT} have explicitly shown 
 that there is a precise correspondence between a subset of the terms 
 in the one loop Matrix theory potential and the linearized DLCQ 
 supergravity potential arising from exchange of quanta with zero 
 longitudinal momentum. 
 These authors also point out, however, that the finite $N$ Matrix model 
 violates the equivalence principle. At finite $N$, then, Matrix theory 
 and DLCQ supergravity are distinct. As a consequence of this, the authors
 of \cite{KT} espouse the view that near DLCQ M-theory
 is not described at low energies by near DLCQ supergravity.
 This possibility has also been taken seriously by Banks \cite{Ba} and 
 Susskind \cite{Sulect}. The latter author suggests that amplitudes computed
 in the finite $N$ Matrix model and near DLCQ supergravity will agree only
 for special processes, presumably protected by a supersymmetry
 nonrenormalization theorem which allows 
 a continuation from the Matrix model to the supergravity regime. 
 
 An alternative resolution has been advocated by Douglas and 
 Ooguri \cite{DOS}. This is that the DLCQ Lagrangian is renormalized in a 
 nontrivial manner when modes with zero longitudinal momentum are 
 integrated out. If this were correct, it should be possible to find a 
 modified Lagrangian which yields an adequate description of the physics. 
 
 Even though the widespread hope remains that the above difficulties
 of the finite $N$ model will be removed as $N\to\infty$, 
 Banks \cite{Ba} has recognized the possibility that the large $N$ limit
 of near DLCQ M-theory might not converge to the eleven dimensional theory.
 
 Susskind \cite{Sulect}, on the other hand, has argued that near DLCQ M-theory
 is, in the large $N$ limit, able to capture eleven dimensional physics,
 in spite of the fact that in the IMF it is manifestly defined as a
 compactification on a zero size spatial circle. His point is simply that
 an object of proper longitudinal size $L$ is Lorentz contracted to
 a size $\propto LR/N$ in a frame where its longitudinal momentum is
 $N/R$ ($R$ being the radius of the spatial circle). Clearly this 
 size can be made arbitrarily smaller than $R$ for large $N$.
 
 Balasubramanian, Gopakumar and Larsen \cite{BGL} have given a 
 very precise discussion of the limits involved in
 the definition of the Matrix model and have
 provided additional evidence for the possibility of recovering
 the full uncompactified theory in the large $N$ limit.
 In particular, they examine a classical solution of eleven
 dimensional supergravity with a compactified longitudinal
 direction of radius $R$, carrying $N$ units of momentum in the 
 compact direction. They discover that, even for $R \to 0$, the physical
 size of the circle, as measured in the supergravity metric, can be 
 made arbitrarily large (for all finite transverse distances) 
 by increasing $N$. This establishes at least the self-consistency of 
 the supergravity solution in this limit. 
 
 On top of this, Bilal \cite{Bi} has analyzed a string theory one-loop
 amplitude in the near DLCQ limit, finding it to have a 
 well-defined, finite limit. This raises the hope that the situation 
 in M-theory might be simpler than that discussed in \cite{HP} for field 
 theory. The crucial difference between the string and field theory
 cases is the existence of string winding modes.
 
 Despite all of this, a skeptic might remain unconvinced. Arguments
 based on Lorentz contraction might work out differently in a 
 theory with extended objects, which can wrap around the compact
 direction. Also, the case might be made that the self-consistency of
 a supergravity solution is logically independent from the recovery
 of the full eleven dimensional M-theory.
 
 The essence of the matter is that there is a potential contradiction 
 in some of the recent attempts to clarify the significance of Matrix 
 theory: on the  one hand, a proof of the validity of the Matrix model
 based on  knowledge gained from perturbative 
 type IIA string theory requires $R$ to be small; on the other hand, 
 for the model to describe uncompactified M-theory it would seem 
 necessary to let $R \to \infty$ \cite{BFSS}. Even if this is done in two 
 separate steps, as in the near DLCQ program \cite{Su,Se} (first let 
 $R\to 0$ at fixed $N$, then take $N,R' \to \infty$), arguments 
 which suggest that the large $N$ Matrix model describes eleven 
 dimensional M-theory \cite{Sulect,BGL} are at risk of being 
 in conflict with Seiberg's proof \cite{Se}, which is based on the 
 understanding that, for any $N$, the Matrix model describes M-theory 
 in the IMF, compactified on a vanishingly small spatial circle.
 Consequently, the question of whether Matrix theory  provides
 information about uncompactified M-theory seems to merit 
 further discussion. That is the general motivation for the present 
 paper.
 
 If we are to believe that a theory which in the IMF is compactified
 on a vanishingly small circle can capture (in the large $N$ limit) 
 the physics of the same theory with no compactification, then it seems
 inevitable that a compatification in the IMF of {\em any finite size}
 should also do so. Any of the decompactification arguments advanced for
 the case of $R=0$ would certainly apply for any other value of $R$. 
 Unless there are several uncompactified limits
 of the theory, one is led to the conclusion that physics
 in the IMF for such a theory must be {\em independent of the 
 compactification radius}. (The authors of \cite{BGL} have also arrived
 at this conjecture.) This is the specific question that we investigate
 in what follows.
 
 We will begin by giving, in Section \ref{boostsec}, 
 a precise specification
 of the limits that concern us. In Section \ref{fieldsec}
 we then attempt to 
 analyze the problem for a theory of point particles; we consider
 scalar field theory as a concrete example.
 In Section \ref{stringsec} we reexamine this issue in a theory with extended 
 objects, string theory. We conclude in Section \ref{Msec}
 with some discussion on the possible implications of our results 
 for the case of M-theory. 
  
\section{Boosting to the Infinite Momentum Frame}
\label{boostsec}

We wish to consider the amplitude for a scattering event in a
(Lorentz invariant) theory defined in $D$ spacetime dimensions.
The situation will be analyzed from the 
vantage point of different Lorentz frames, related  by 
boosts along what we will call the longitudinal direction. 
Specifically, we introduce a frame \Fp, 
and a family of frames \Fe, indexed by a parameter $\eps\ll 1$,
and related to \Fp\ by a boost with rapidity
parameter $\omega=\ln(\sqrt{2}/\eps)$, i.e. $\beta = \tanh \omega$ 
is given by \footnote{Notice that our
definition of $\beta$ in terms of \eps\ coincides with that of \cite{Bi},  
and is slightly different from the one in \cite{Se}. Both agree as
$\eps\to 0$.}  $\beta \approx 1-\eps^{2}$.
As $\eps\to 0$, \Fe\ approaches the IMF. By this we simply mean
that all longitudinal momenta in \Fe\ become arbitrarily larger than 
any other scale in the problem. With this understanding, we
will loosely refer to \Fe\ as the IMF.
If the theory is quantized on an equal time surface in \Fe, then as
$\eps\to 0$ the quantization surface approaches a light 
front surface in \Fp. Because of this we will refer to \Fp\
as the light front quantization (LFQ) frame.

Notice that there is in 
our discussion a clear distinction between the IMF and the LFQ
frame, whereas both terms are sometimes used interchangeably in the 
literature. In particular, the momenta of localized objects
involved in the scattering process are fixed and finite in the 
LFQ frame \Fp. Of course, the same 
physics should be visible from the point of view of either frame 
(even though, as a result of the compactification which will be
introduced below, longitudinal boosts are not a symmetry of the theory).

Following Seiberg \cite{Se}, we compactify the longitudinal direction 
in \Fe, $x^{1}$, on a circle of radius $R$,
\be
{x^{0} \choose x^{1}} \sim {x^{0} \choose x^{1}} + 2\pi R{ 0\choose
 1}.
\ee
In terms of \Fp\ light front coordinates 
$x'^{\pm}= (x'^{0}\pm x'^{1})/\sqrt{2}$ this is
\be \label{nearDLCQ}
{\xp \choose \xm} \sim {\xp \choose \xm} + 2\pi R'
{ \eps^{2}/2 \choose -1},
\ee
where we have defined $R'=R/\eps$. In \Fp, then, the compactification
is almost along $\xm$, with radius $R'$. 

The longitudinal momentum in \Fe\ is quantized; let $P=N/R$ be the 
total initial longitudinal momentum for the scattering event.
In \Fp\ this corresponds to the statement that the generator of
$\xm$ translations (at fixed $\xp$) is quantized (up to 
$\calO(\eps^{2})$ terms\footnote{Alternatively \cite{HP,Bi}, one can 
introduce $t'=\xp+\eps^{2}\xm/2=\eps x^{0}$; the periodic identification is
then at fixed $t'$.  If $\Pm$ is defined as the generator of $\xm$ 
translations at fixed $t'$, then it is exactly quantized, $\Pm=-N/R'$.
One must bear in mind, however, that the metric in the  $(t',\xm)$
coordinates is nontrivial.}),
$\Pup=-\Pm=N/R'$. The situation is summarized in Fig.~1.

\begin{figure}[ht]  \label{frames}
 \begin{center} 
  \setlength{\unitlength}{1cm}
  \begin{picture}(8,1.3)
   \put(1.0,0.0){\shortstack[c]{\Fe\\$R$\\$P=N/R$}}
   \put(5.0,0.0){\shortstack[c]{\Fp\\$R'=R/\eps$\\$\Pup=N/R'$}}
   \put(4.8,0.9){\vector(-1,0){1.9}}
   \put(3.3,0.6){$\scriptstyle e^{\omega}=\sqrt{2}/\eps$}
  \end{picture}
  {\small \caption{Relation between the IMF \Fe, and the LFQ frame
    \Fp. The arrow indicates a longitudinal boost.}}
 \end{center}
\end{figure}

We should emphasize that the compactification depends on \eps\ not 
only because the proper length of the circle, $2\pi R$, might be 
$\eps$-dependent (see below), but also because the frame \Fe\ where
the circle is purely spatial is different for different values of \eps.

Now, by construction, $\eps \to 0$ is the IMF (or LFQ) limit.
There are, however, several
ways of taking this limit. We will be interested in the
following three:

\newcounter{count}
\begin{list} {\arabic{count}.}{\usecounter{count}}
 \item {\bf BFSS (uncompactified IMF) limit:} $\eps\nb\to\nb 0$ with 
 $N\nb\to\nb\infty$, $R\nb\to\nb\infty$, $N/R\nb\to\nb\infty$.
 Then $R'=\eps^{-1}R \to \infty$.
 
 \item {\bf Compactified IMF limit:} $\eps\nb\to\nb 0$ 
  with $N\nb\to\nb\infty$,
  $R$ fixed. Then $R'=\eps^{-1}R\nb\to\nb\infty$.
  
 \item {\bf Seiberg-Susskind (near DLCQ) limit:} 
  $\eps\nb\to\nb 0$ with $N,R'$ fixed\footnote
  {Seiberg's construction \cite{Se} involves also an 
  $\eps$-dependent change of units in \Fe. This is irrelevant for the 
  present discussion. So, to avoid misunderstandings,
  we emphasize that until 
  Section \ref{Msec} we will use the same units in \Fe\ and \Fp.}.
  Then $R=\eps R'\nb\to 0$, $N/R=\eps^{-1}N/R'\nb\to\nb\infty$.
\end{list}

Notice that, from (\ref{nearDLCQ}), we need $R'\eps^{2} \to 0$ in 
order for the compactification in \Fp\ to converge to one at fixed 
$\xp$. This gives for limit 1 the additional requirement
$R\propto\eps^{-\alpha}$ with $0<\alpha<1$, and therefore 
$N\propto\eps^{-\beta}$, with $\beta>\alpha$. We should perhaps stress 
that in limits 1 and 2 the circle does not become null as $\eps\to 0$.

In the first two limits, the behavior of $N/R'$ depends on the
nature of the scattered objects. The simplest case is that with 
objects which are localized in the longitudinal direction.
Then $N/R'$ should be fixed and finite.
This implies $N/R \propto \eps^{-1}$ in limit 1 
(i.e. $\beta=\alpha + 1$), and $N \propto \eps^{-1}$ in limit 2.

For an extended longitudinally wrapped object, on the other hand, it 
is not the momentum $N/R'$, but the momentum density, $N/R'^{2}$, that
should be fixed and finite. For limits 1 and 2 this means 
$N/R \propto \eps^{-2}$ ($\beta=\alpha + 2$) and 
$N \propto \eps^{-2}$, respectively.
(To be absolutely clear we remark that the above relations between
$N$ and $\eps$ give only the leading small $\eps$ behavior.)

In the case of limit 3,  we imagine that $R' \to \infty$  {\em after} 
$\eps\to 0$. Then $N \to \infty$, again with two possible 
choices for the behavior of $N/R'$. 

It is evident that limit 1 yields the uncompactified theory. 
The situation is not so clear for limit 2. There $R' \to \infty$, 
but $R$ is fixed. Which is it that is relevant? Things
are even worse for limit 3. Even if we let $R'\to\infty$, we have taken 
$R \to 0$ to begin with. So again, which is the physically
significant radius?

\begin{figure}[ht] \label{decompact}                 
 \begin{center}
 \leavevmode
 \epsfbox{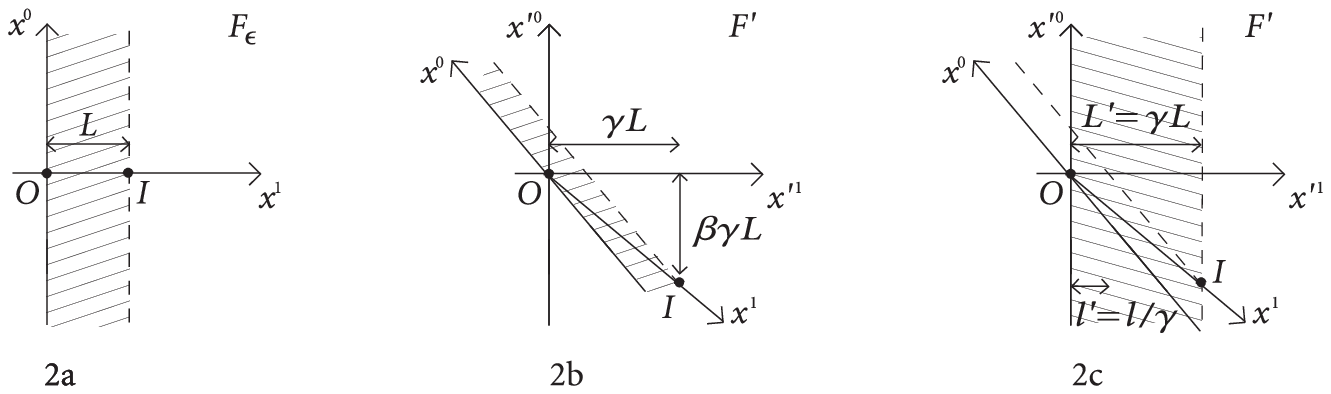}
 \end{center}
 \caption[]{Decompactifying by boosting?}
\end{figure}

To try to answer this question, consider Fig.~2.
Fig.~2a shows the $(x^{0},x^{1})$ plane in \Fe, 
where the periodic 
identification is at constant $x^{0}$. So, for example, the origin 
$O$ and the event $I$, a distance $L=2\pi R$ away, are identified.
The shaded area is a physical 
slice of the covering space; all events in the compactified spacetime 
are included in this slice exactly once. Fig.~2b shows events $O$ 
and $I$, and the same physical slice, in \Fp. The periodic 
identification now involves shifts by $\gamma L$ along $x'^{1}$, and 
by $-\gamma \beta L$ along $x'^{0}$. Fig.~2c also shows \Fp,
but with a different, equally valid, choice of physical slice,
one which is parallel to the $x'^{0}$ axis.
From the figure it is easy to convince oneself 
that this slice also contains all physical events exactly once. To 
avoid confusion, we emphasize that the periodic identification in this 
slice is still made diagonally, not horizontally.

Now, the point is that in Fig.~2c the `spatial extent' of the 
slice as seen in \Fp\ is $L' = \gamma L$, which can be made
arbitrarily large. Since $\gamma = \cosh \omega
\approx \eps^{-1}/\sqrt{2} \to \infty$, this is precisely what 
allows $L' \to \infty$ for limit 2, and $L'$ fixed in spite of $L \to 
0$ in limit 3. 

Notice that this seems to be the {\em opposite}\/ of Lorentz
contraction. That is to say, if an object of proper length $l$ just slightly
shorter than $L$ were to be placed at rest along the $x^{1}$-axis in \Fe,
with its left end at the origin, then its length in \Fp\ would be
$l' = l/\gamma$, a factor $\gamma^{2}$ shorter than $L'$ (see 
Fig.~2c).
But by the very definition of the IMF, there is no object at rest 
in \Fe, so this should be stated the other way around: an object of 
proper length $s'\ll L'$ at rest in \Fp\ has length $s=s'/\gamma \ll L$ 
in \Fe. (This is essentially Susskind's Lorentz contraction 
argument \cite{Sulect}.)

Something very perplexing is going on here. On the one hand, Fig.~2 
seems to indicate that a space can be decompactified by a large boost.
On the other hand, the proper length of the circle, $L=2\pi R$, is of
course Lorentz invariant.

Let us now suggest that, in a point particle theory, the relevant 
criterion for decompactification should be that the Compton wavelengths
of all particles be, in the IMF,
much smaller than the spatial radius $R$.\footnote{We thank Sanjaye 
Ramgoolam for suggesting this.}
But by the  above
Lorentz contraction argument this is accomplished automatically as 
$N\to\infty$, independently of $R$. It is then plausible to expect 
scattering amplitudes in a point particle theory
to be independent of $R$ in this
limit. We explore this possibility in Section \ref{fieldsec}.
We stress that this expectation is not in any
obvious way related to the well-known irrelevance of the  
(frame-dependent) radius of the null circle in DLCQ amplitudes. 
$R$\/ in our discussion is defined in \Fe\ as a purely spatial radius.

In a theory with extended objects, on the other hand, the proper length 
of the circle could possibly be `felt' by longitudinally wrapped 
objects. In particular, the mass of such an object should be 
$R$-dependent. One might then expect amplitudes in such a theory to 
depend on $R$\/ even in the large $N$ limit. 
We will examine this in Section \ref{stringsec}.

\section{Field Theory Amplitudes} \label{fieldsec}

We now study the $R$-dependence of $D$ dimensional field theory scattering 
amplitudes in the IMF \Fe. For concreteness, we will focus on the case of 
a scalar field; our arguments should be easy to generalize.

Consider a scattering process with $A$ initial and $B$ final particles
of  momenta $\{K^{M}_{a}\}$. Here $a=1,\ldots, A+B$ is a particle label,
and $M=0,1,\ldots,D-1$  a coordinate index. 
We split the external momenta into longitudinal and transverse 
parts according to $K^{M}_{a}=(K^{0}_{a},K^{1}_{a},K^{i}_{a})=
(E_{a},P_{a},\vec{Q}_{a})$, where 
$i=2,\ldots,D-1$ (and $x^{1}$ is the compact direction). We take all 
$E_{a}> 0$; in other words, the initial (final) external legs are 
incoming (outgoing). Due to the 
compactification, all longitudinal momenta are quantized, $P_{a}=N_{a}/R$.
By the definition of the IMF, $N_{a}>0 \; \forall\; a$.
Let $N= \sum_{a=1}^{A} N_{a}$; then $P=N/R$ is the total (initial)
longitudinal momentum.
We will examine the compactified IMF limit $\eps \to 0$ with $N \to \infty$,
$R$ fixed (and therefore $P \to \infty$) of the amplitude. The fraction of 
longitudinal momentum of each external leg, $\zeta_{a}=N_{a}/N$, is 
held fixed as $N \to \infty$.

The first thing to notice is that kinematic Lorentz invariants 
$K_{a}\cdot K_{b}$ could just as well be evaluated in \Fp,\/ where they 
are finite. So when evaluated in \Fe, to leading order in $N$ they
can only depend on ratios of 
longitudinal momenta, and are therefore independent of $R$.  This 
already guarantees that all tree level diagrams in the IMF do not 
depend on $R$.
 
The situation could appear to be different for loop diagrams, which
include sums over the longitudinal momenta carried by intermediate 
lines, weighted by explicit factors of $1/R$. As usual in IMF (or LFQ)
physics, things are more transparent in old-fashioned perturbation theory
(OFPT) language \cite{Wein}.

\begin{figure}[ht] \label{OFPT}                  
 \begin{center}
 \leavevmode
 \epsfbox{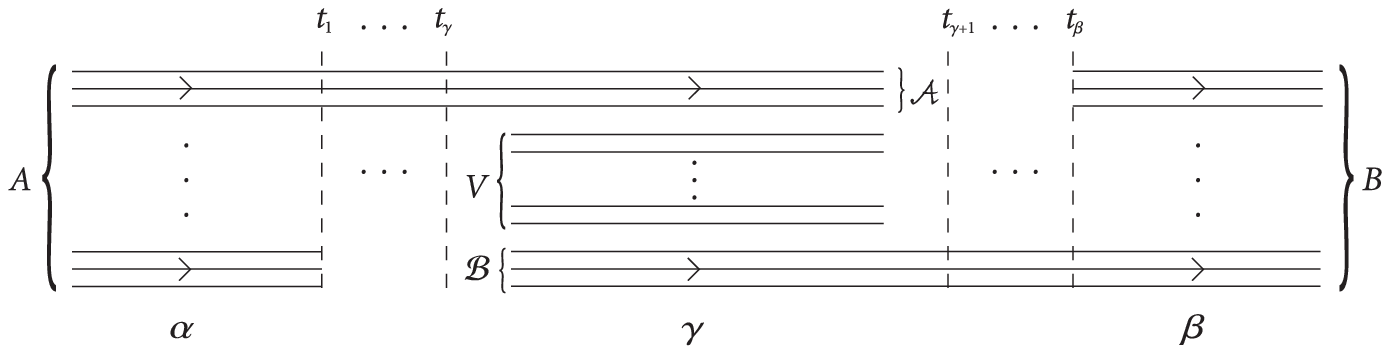}
 \end{center}
 \caption[]{Arbitrary old-fashioned perturbation theory diagram. 
  $\alpha$, $\beta$, and $\gamma$ denote the initial, final and
  intermediate state, respectively.}
\end{figure}

Consider then the arbitrary OFPT diagram shown in Fig.~\ref{OFPT}.
In some intermediate state (between two interaction times 
$t_{\gamma}$, $t_{\gamma +1}$)
the diagram has $I$ particle lines, of which \calA\
(\calB)\ are external initial (final) lines. Relabel these 
so that they have momenta
$K^{M}_{a}$, $a=A-\calA+1,\ldots,A$ ($a=A+1,\ldots,A+\calB$).
The remaining
$V=I-\calA -\calB$ lines of the
intermediate state correspond to internal (i.e. virtual) 
particles with momenta  
$k^{M}_{a}=(e_{a},p_{a},\vec{q}_{a})$,\ $a=1,\ldots,V$. Here 
$p_{a}=n_{a}/R$; define the 
longitudinal fraction $\eta_{a}=n_{a}/N$. Notice $n_{a}$ and 
$\eta_{a}$ are not necessarily positive. 
The energy difference between the initial state $\alpha$ and an
intermediate state $\gamma$
with a particlular choice of $\eta_{a}$ is
\be
E_{\alpha}-E_{\gamma}=
 \sum\limits_{a=1}^{A}
   \sqrt{\zeta_{a}^{2}P^{2} + \vec{Q}_{a}^{2}}
 -\sum\limits_{a=A-\calA+1}^{A+\calB}\sqrt{\zeta_{a}^{2}P^{2}
  + \vec{Q}_{a}^{2}}
 -\sum\limits_{a=1}^{V}\sqrt{\eta_{a}^{2}P^{2}
  + \vec{q}_{a}^{\,2}}.
\ee
Such a state would thus contribute to the amplitude a 
factor\footnote{The normalization factors $1/2e_{a}$ are needed to 
reconstruct the covariant scalar field propagators.}

\bea \label{f}
f(P,\eta_{a}) &=&
 \prod_{a=1}^{V}\left(\frac{1}{2\sqrt{\eta_{a}^{2}P^{2}
 + \vec{q}_{a}^{\,2}}}\right)  \\
{}&{}& \times \;
\frac{1}{
 \sum\limits_{a=1}^{A-\calA}
   \sqrt{\zeta_{a}^{2}P^{2}+ \vec{Q}_{a}^{2}}
 -\sum\limits_{a=A+1}^{A+\calB}
   \sqrt{\zeta_{a}^{2}P^{2}+ \vec{Q}_{a}^{2}}
 -\sum\limits_{a=1}^{V}\sqrt{\eta_{a}^{2}P^{2}
  + \vec{q}_{a}^{\,2}}}. \nonumber
\eea

The total factor associated with the state is therefore (omitting the 
integrals over the transverse momenta $\vec{q}_{a}$)

\be \label{F}
F_{N}(R) = 
 \left(\frac{1}{R}\right)^{V}\sum_{n_{1}=-\infty}^{+\infty}
   \cdots  \sum_{n_{V}=-\infty}^{+\infty} R\  
 \delta_{N_{\alpha},N_{\gamma}}
\ f\left(\frac{N}{R}, \frac{n_{a}}{N}\right),
\ee
where the Kronecker delta enforces longitudinal momentum conservation,
\be \label{Ncons}
N_{\alpha}-N_{\gamma} =
 \sum\limits_{a=1}^{A-\calA}N_{a}
  -\sum\limits_{a=A+1}^{A+\calB}N_{a}
  -\sum\limits_{a=1}^{V}n_{a}
  =0.
\ee

Now, as $N \to \infty$ with $N_{a}\propto N$,  (\ref{Ncons}) shows that 
in the generic case at least one of the $n_{a}$ must scale like $N$.
An exception to this occurs if it so happens that
$\sum_{a=1}^{A-\calA}N_{a}=\sum_{a=A+1}^{A+\calB}N_{a}$. 
Let us discuss the generic situation for now;
we will return to the exceptional case below.
  
Any one term in the sum (\ref{F}), with all (but one)  $n_{a}$ fixed,
vanishes
in the large $N$ limit. To properly examine the limit, then,
focus attention
on terms in which all $n_{a}$ scale like $N$ (all $\eta_{a}$ are then
held fixed).
It is straightforward to verify that   
the dominant contribution to $F_{N}(R)$ comes from those states with 
$n_{a}\ge 0 \ \forall \ a=1,\ldots,V$. For these
the leading terms in the energy 
difference denominator in (\ref{f}) cancel.
States with negative $n_{a}$ are 
suppressed by an additional factor of $(N/R)^{-2}$. This is just the usual 
decoupling of negative momentum states in the IMF \cite{Wein}. Omitting these, 
rewriting the sums in terms of $\eta_{a}$, and replacing $R$ by $N 
\cdot R/N$,
we have
\be \label{simplF}
F_{N}(R) =\left(\frac{1}{N}\right)^{V}
\sum_{\eta_{1}=0,\frac{1}{N},\ldots}^{1}\cdots
\sum_{\eta_{V}=0,\frac{1}{N},\ldots}^{1} \left(\frac{N}{R}
\right)^{V-1}  N \delta_{N_{\alpha},N_{\gamma}} \
f\left(\frac{N}{R},\eta_{a}\right).
\ee
The point is now simply that as $N \to \infty$, the sums
converge to integrals of a finite integrand. To make this clear,
define a sequence of step-functions $\{ g_{N}(R,\eta_{a}) \} $ by
\be \label{g}
{g_{N}(R,\eta_{a}) =\left(\frac{N}{R}\right)^{V-1}
 f\left(\frac{N}{R},\frac{\left[ N \eta_{a} 
\right]}{N} \right),}
\ee
where $[ N \eta_{a} ]$ denotes the integral part of $N\eta_{a}$. Then 
(\ref{simplF}) can be written as an integral,

\be \label{intF}
F_{N}(R) =
\int\limits_{0}^{1+1/N}d\eta_{1}\cdots \int\limits_{0}^{1+1/N}d\eta_{V} 
\  N \delta_{N_{\alpha},N_{\gamma}} \ g_{N}(R,\eta_{a}).
\ee

As $N \to \infty$, the sequence $g_{N}(R,\eta_{a})$ converges 
almost everywhere to the function

\be \label{ginfty}
{g_{\infty}(\eta_{a}) =\prod_{a=1}^{V}\left(\frac{1}{2\eta_{a}}\right)
\frac{2}
 {\sum\limits_{a=1}^{A-\calA}\zeta_{a}^{-1}\vec{Q}_{a}^{2}
 -\sum\limits_{a=A+1}^{A+\calB}\zeta_{a}^{-1}\vec{Q}_{a}^{2}
 -\sum\limits_{a=1}^{V} \eta_{a}^{-1} \vec{q}_{a}^{\,2}}},
\ee
and $ N \delta_{N_{\alpha},N_{\gamma}}$ becomes a delta function.
Consequently, $F_{N}(R)$ converges to the {\em $R$-independent}\/ factor
\be \label{Finfty}
{F_{\infty} =
\int\limits_{0}^{1}d\eta_{1}\cdots \int\limits_{0}^{1}d\eta_{V} 
\ \delta(\sum\limits_{a=1}^{A-\calA}\zeta_{a}
  -\sum\limits_{a=A+1}^{A+\calB}\zeta_{a}
  -\sum\limits_{a=1}^{V}\eta_{a}) \  g_{\infty}(\eta_{a})} 
.
\ee
We have ignored here possible 
subtleties in exchanging the limit and the integrals, as well as in
regularizing the integrals. Nonetheless, the final result is eminently 
reasonable, since it amounts to the usual cancellation of factors of 
$P$ in IMF diagrams that yields finite perturbation theory rules
(matching those of LFQ), coupled with the standard conversion of sums
into integrals in the large $N$ limit.

We have thus shown that {\em generic} IMF
scattering amplitudes in a point particle theory are independent of 
the IMF compactification radius, as conjectured in Section \ref{boostsec}.
To be precise, we emphasize that we 
are discussing here the compactified IMF limit
$N \to \infty$ with $R$ fixed. The large $N$ limit is essential;
it has the double effect of forcing $P \to \infty$ and turning the 
sums into integrals. The essence of the matter is very simple: in this 
limit, the amplitudes are functions of $R$
only through the
combination $N/R$, so all dependence on $R$ drops out for large $N$.

It is important to note that expression (\ref{Finfty}) agrees precisely 
with the intermediate state factor that would be obtained with 
standard IMF (or LFQ) perturbation theory rules \cite{Wein}
in the {\em uncompactified} theory.

The situation is quite different, however, in the exceptional case 
where the external lines which appear
in the intermediate state are such that
$\sum_{a=1}^{A-\calA}N_{a}=\sum_{a=A+1}^{A+\calB}N_{a}$. For this 
case, as can be seen from (\ref{Ncons}), the kinematics forces the total
internal momentum to vanish,
$\sum_{a=1}^{V}n_{a}=0$, and the $n_{a}$ need not scale with $N$. Any 
one term in the sum (\ref{F}), with all $n_{a}$ fixed, converges to a 
{\em finite} value as $N \to \infty$,
\be
\left(\frac{1}{R}\right)^{V-1}
f\left(\frac{N}{R}, \frac{n_{a}}{N}\right) \to
\prod_{a=1}^{V}\left(\frac{1}{2\sqrt{\left(\frac{n_{a}}{R}\right)^{2}
    + \vec{q}_{a}^{\,2}}}\right) 
\frac{1}{-\sum\limits_{a=1}^{V}
   \sqrt{\left(\frac{n_{a}}{R}\right)^{2}+ \vec{q}_{a}^{\,2}}}.
\ee
In this case, then, lines with negative longitudinal momentum do not
decouple\footnote{This possibility was ignored in \cite{Wein}.},
and there is some remaining $R$-dependence even for $N 
\to \infty$.

A particular example of this exceptional case is a
diagram recently examined by Hellerman and Polchinski \cite{HP}.
They consider a $2 \to 2$ one-loop covariant diagram
with no longitudinal momentum exchange,
in $\phi^{4}$-theory, and find that it diverges in the Seiberg-Susskind
limit, $R \to 0$ at fixed $N$.\/ This can of course be seen directly in 
OFPT language.

\begin{figure}[ht]                               
\begin{center}
\leavevmode
\epsfbox{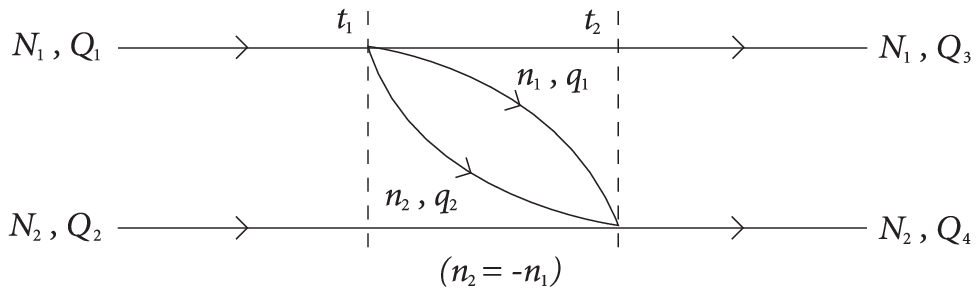}
\end{center}
\caption[]{An `exceptional' OFPT one-loop diagram.}
\end{figure}

Fig.~4 shows a particular time ordering of the one-loop covariant
diagram in \cite{HP}. In our notation (using $n_{2}=-n_{1}$),
the intermediate state contributes a total factor
\bea \label{hpF}
F_{N}(R) &=&
 \frac{1}{R} \sum_{n_{1}=-\infty}^{\infty}
 \frac{1}{2\sqrt{\left(\frac{n_{1}}{R}\right)^{2}+\vec{q}_{1}^{\,2}}}
 \, 
 \frac{1}{2\sqrt{\left(\frac{n_{1}}{R}\right)^{2}+\vec{q}_{2}^{\,2}}}
   \\
{}&{}& \times\:\frac{1}{
  \sqrt{\left(\frac{N_{1}}{R}\right)^{2}+\vec{Q}_{1}^{2}}
 -\sqrt{\left(\frac{N_{1}}{R}\right)^{2}+\vec{Q}_{3}^{2}}
 -\sqrt{\left(\frac{n_{1}}{R}\right)^{2}+\vec{q}_{1}^{\,2}}
 -\sqrt{\left(\frac{n_{1}}{R}\right)^{2}+\vec{q}_{2}^{\,2}}
     }. \nonumber
\eea

As $R \to 0$ with all $N_{a}$ fixed, all terms in the sum with 
$n_{1}\neq 0$ vanish ($\propto R^{2}$), but the zero mode
term with $n_{1}=0$ diverges ($\propto R^{-1}$). This is what was found 
in \cite{HP}. 
In the compactified IMF limit $N \to \infty$ with $R$ fixed (and 
$N_{a}\propto N$), on the other hand, each term in the sum converges 
to a finite limit, as stated before. $F_{\infty}(R)$ itself is 
then finite (for $R>0$) and $R$-dependent.
Physically, the kinematics is such that, even in the IMF,
virtual states are allowed to have particle lines with small 
longitudinal momentum. These particles have arbitrarily
large Compton wavelengths, and can therefore `feel' the size of the 
compactification circle.

To summarize, we have shown that most field theory amplitudes
are independent of $R$ in the compactified IMF limit. There are
exceptional processes, however, which remain $R$-dependent even in the 
large $N$ limit: those with zero longitudinal momentum transfer between
two subsets of the initial and final particles. Intermediate states in
OFPT diagrams for such processes can have internal lines with
arbitrary (not necessarily large) longitudinal momenta, corresponding
to virtual particles which can detect the finite size of the 
compactification. Hence, we seem forced to conclude then that the truly 
decompactified theory (for all possible scattering events) 
is only obtained as $R \to \infty$.

\section{String Theory Amplitudes} \label{stringsec}

We now examine the $R$-dependence of superstring amplitudes in the 
IMF \Fe.  A natural question to consider is whether this dependence
is in any way constrained by the general properties of string theories. 
T-duality, in particular, implies an equivalence between theories at 
different radii, and so appears to be relevant to the present analysis. 

To examine this more closely, consider for concreteness type IIA string 
theory with coupling \gs\ and string length \ls, compactified to nine 
dimensions on a circle of radius $R$.
The crucial point to realize is that, while it is true that a 
scattering process in this theory is equivalent under T-duality to  
a process in a space with compactification radius $\hat{R}=\ls^{2}/R$, 
this dual process takes place in a {\em different} theory, namely 
type IIB string theory with coupling $\hat{g}_{s}=\gs (\ls/R)$.
(In addition, the external longitudinal momentum and winding 
numbers are interchanged.)  This is just the point that T-duality 
expresses not so much a symmetry of either theory as a relation 
between the two theories, or better yet, a translation between two 
different descriptions of the same physics. 

While T-duality does not directly imply a connection between two 
different values of $R$ for a given theory, 
the passage to a dual description can of course provide insight on 
the nature of a specific limit of the theory. For instance, as a 
result of T-duality, a string theory compactified on a 
vanishingly small ($R\to 0$) circle is still able to fully capture the 
behavior of an uncompactified theory ($\hat{R}\to\infty$). Notice, 
however, that this alone does not immediately guarantee that string 
amplitudes are well-behaved in this (the Seiberg-Susskind) 
limit \cite{Bi}, for one must bear in mind that the dual amplitudes 
are only known to be finite when expressed in terms of $\hat{g}_{s}$. 
Since \gs\ is held fixed as $R\to 0$, the dual coupling 
$\hat{g}_{s}=\gs (\ls/R)\to\infty$.\footnote{One can switch to an
S-dual description of the type IIB theory, with dual coupling 
$\bar{g}_{s}=\hat{g}_{s}^{-1}\to 0$. In this language one is considering 
the amplitude for scattering D-strings in an uncompactified theory 
with string length $\bar{l}_{s}=\hat{g}_{s}^{1/2}\ls\to\infty$, which
is still potentially singular.} Thus, to determine the behavior of the 
amplitudes in the Seiberg-Susskind limit, one cannot escape the need to 
carry out an explicit calculation. 
This is even more evident in the compactified IMF limit 
which concerns us here, where $R$ is held fixed while $N\to\infty$.

We now focus attention on a concrete example: a one-loop amplitude
in type II string theory. The simplest non-trivial case is that with 
four bosonic vertex operators. This (in the special case with no external
winding) was recently considered by Bilal \cite{Bi}, with the 
purpose of studying the behavior of the amplitude in the
Seiberg-Susskind limit, $R \to 0$ at fixed $N$, 
and comparing it against the results of 
Hellerman and Polchinski \cite{HP}.

The amplitude is computed in the Green-Schwarz light cone formalism; 
our notation is as in \cite{GSW,Bi}.  The GS light cone (which is 
unrelated to the LFQ in \Fp) is taken along direction $M=9$.
As before, the compact direction is $M=1$. Coordinate indices are 
split accordingly into $M=(0,i,9)$, with $i=1,\ldots,8$, or 
$M=(1,\mu)$, with $\mu=0,2,\ldots,9$.
In addition, subindices $a,b=1,\ldots,4$ label the external vertices.

One must compute
\bea \label{Adef}
A_{\mathrm{cl}}^{(4)} &=&
   \left(\frac{\kappa}{4\pi}\right)^{4}\int\frac{d^{2}w}{|w|^{2}} 
   \int\prod\limits_{a=1}^{3}\frac{d^{2}\rho_{a}}{|\rho_{a}|^{2}} \ I,
   \mbox{\qquad where}\\
I &=&\frac{1}{R}\sum\limits_{n,m=-\infty}^{+\infty}
   \int d^{9}k^{\mu}\ 
   \mbox{Tr}\left[V(k_{1},\rho_{1})\cdots V(k_{4},\rho_{4})
   w^{L_{0}}\bar w^{\tilde L_{0}}\right], \nonumber 
\eea
$\rho_{a}=z_{1}\cdots z_{a}$, and $w=\rho_{4}$. We have let 
$k^{\mu}$,  $n$ and $m$ above stand for the transverse momentum,
longitudinal momentum number, and longitudinal winding number running
around the loop, respectively.

The right- and left-moving momenta are defined as 
\be \label{kR,L}
k_{a R}^{\mu}=k_{a L}^{\mu}=k_{a}^{\mu},\qquad k_{a R}^{1}=
\frac{n_{a}}{R} - \frac{m _{a} R}{\ap},\qquad  k_{a L}^{1}=
\frac{n_{a}}{R} + \frac{m _{a} R}{\ap}.
\ee
For now, restrict attention to the scattering of four bosonic
ground states with no external 
winding, $m_{a}=0 \ \forall \ a$. 
As explained in \cite{GSW}, because of the fermionic zero mode trace
the vertex operators can be effectively taken to be of the form 
\be \label{vertex}
V(k,z)= \frac{1}{4}\varepsilon_{il}
\left[R_{0}^{ij}k_{R}^{j}
\exp\left(ik_{R}\cdot X_{R}(z)\right)\right]
\left[\tilde R_{0}^{lj}k_{L}^{j}
\exp\left(ik_{L}\cdot X_{L}(\bar z)\right)\right],
\ee
where $R_{0}^{ij}=S_{0}^{\alpha}\gamma_{\alpha \beta}^{i 
j}S_{0}^{\beta}/4$ (and similarly for $\tilde R_{0}^{ij}$).
The detailed calculation is given in  \cite{Bi}. Using the standard 
torus coordinates $\nu_{a}=\ln(\rho_{a})/2\pi i$, and modular parameter
$\tau=\ln(w)/2\pi i$, the amplitude can (after a Poisson resummation on 
the winding number $m$) be put in the form
\bea \label{Anowind}
A_{\mathrm{cl}}^{(4)} &=& \frac{(\kappa \pi)^{4}}{\ap^{5}}
  K_{\mathrm{cl}}
  \int\frac{d^{2}\tau}{(\mathrm{Im}\tau)^{2}}
  \int\prod\limits_{a=1}^{3}\frac{d^{2}\nu_{a}}{\mathrm{Im}\tau} 
  \ \calI \\
\calI &=& \prod\limits_{a,b}
  \chi(\nu_{ab},\tau)^{\ap k_{a} \cdot k_{b}/2}
    \,\calS   \nonumber\\
\calS &=& \frac{\ap}{R^{2}}\sum\limits_{n,m}
  \exp\left\{-\frac{\pi \ap}{R^{2}\mathrm{Im}\tau}
  \left| m+n\tau +\sum\limits_{a}n_{a}\nu_{a}\right|^{2}
  \right\}. \nonumber
\eea 
Here $\chi(\nu,\tau)=2\pi \exp\{ -\pi 
(\mathrm{Im}\nu)^{2}/\mathrm{Im}\tau \} |E(\nu,\tau)|$,  
$E(\nu,\tau) = {\theta_{1}(\nu|\tau)}/{\theta'_{1}(0 |\tau)}$, 
and $\nu_{ab}=\nu_{a}-\nu_{b}$. 
$K_{\mathrm{cl}}$ is a kinematic factor arising from the trace 
over the fermionic non-zero modes, and is given in \cite{GSW}. 

Using the well-known properties of $\theta_{1}(\nu|\tau)$, it is easy to 
check that \calI\ is invariant under shifts of the insertion 
points $\nu_{a}\to \nu_{a}+1$,\ $\nu_{a}\to \nu_{a}+\tau$ (so that 
$\nu_{a}$ lives on the world-sheet torus defined by $\tau$), and 
under modular transformations (so that $\tau$ is indeed the modular 
parameter, to be integrated over the usual fundamental domain).

We will now examine this amplitude in the compactified IMF limit. For 
concreteness, we regard the amplitude as giving a $2 \to 2$\ scattering 
process, with incoming (outgoing) legs $a=1,2$ ($a=3,4$). Since the 
momenta in the string calculation are defined to satisfy 
$\sum_{a=1}^{4}k_{a}^{\mu}=0$,\ $\sum_{a=1}^{4}n_{a}=0$,\ 
$\sum_{a=1}^{4}m_{a}=0$,
this means that $k_{1,2}^{0}>0$,\ $k_{3,4}^{0}<0$, and therefore (by the 
definition of the IMF) $n_{1,2}>0$,\ $n_{3,4}<0$. Set $N=n_{1}+n_{2}$.
We are interested in $N\to \infty$ with $R$ fixed. 
As in the field theory case discussed in 
Section \ref{fieldsec}, the longitudinal fractions $\eta_{a}=n_{a}/N$ are held 
constant in the limit.
 
The Lorentz invariant factors $k_{a} \cdot k_{b}$ must be finite 
(because they are finite in \Fp ) and 
therefore (as argued in the field theory case) cannot depend 
on $R$.\footnote{For the same reason, the 
corresponding tree amplitude is also guaranteed to be 
$R$-independent.}  The relevant piece of the amplitude 
is thus the sum \calS. Clearly any one term in the sum (with fixed
$n,m$) is exponentially suppressed as $N \to \infty$. To properly
examine the limit, then, we should focus attention on
terms where both $n$ and $m$ scale like $N$. 
Rewriting \calS\ in terms of $\eta=n/N$ and $\zeta=m/N$, we have
\be
\calS = \frac{1}{N^{2}}\sum\limits_{\eta,\zeta}
  \frac{\ap N^{2}}{R^{2}}
  \exp\left\{-\frac{\ap N^{2}}{R^{2}}\frac{\pi}{\mathrm{Im}\tau}
  \left| \zeta+\eta\tau +\sum\limits_{a}\eta_{a}\nu_{a}\right|^{2}
  \right\}.
\ee
Since the spacing of the sums is $1/N$, they converge to integrals 
in the large $N$ limit, just like in the (generic) field theory case. 
Letting $\delta=R/N\sqrt{\ap}$, we see the integrand is of the form
\be
\frac{1}{\delta^{2}}
  \exp\left\{-\frac{1}{\delta^{2}}\frac{\pi}{\mathrm{Im}\tau}
  \left| \zeta+\eta\tau +\sum\limits_{a}\eta_{a}\nu_{a}\right|^{2}
  \right\},
\ee
with $\delta \to 0$. This is precisely the same expression
that had to be considered (for the Seiberg-Susskind limit) in \cite{Bi};
as explained there, the expression converges to the complex delta function
\be
\mathrm{Im}\tau \ \delta^{(2)}\left(\zeta+\eta\tau+\sum _{a} 
n_{a}\nu_{a}\right).
\ee
We conclude that, as $N \to \infty$, the string amplitude without 
external winding becomes $R$-independent, having the form (\ref{Anowind}) 
but with
\be \label{simplS}
\calS= \int d\eta \int d\zeta  \,
\mathrm{Im}\tau \ \delta^{(2)}\left(\zeta+\eta\tau+\sum_{a} 
n_{a}\nu_{a}\right).
\ee
The complex delta function in (\ref{simplS}) can be used to dispose of 
the integrals over $\eta$ and $\zeta$. This gives the simple
result $\calS=1$. The amplitude given by (\ref{Anowind}) is 
then explicitly seen to agree, in the compactified IMF limit,
with that of the {\em uncompactified} ten dimensional theory \cite{GSW}.
 
This is just as for the generic
field theory amplitudes examined in Section \ref{fieldsec}.
Again, the basic idea is that, in the limit of interest, the 
amplitude depends on $R$ only through the combination $N/R$, and so 
all $R$-dependence disappears as $N \to \infty$. This result appears 
therefore to be a general property of string amplitudes (at least those 
with no external winding), not just the one-loop amplitude
examined above.

Notice that, unlike the field theory case, the above
result does not require any special assumptions
about the kinematics of the scattering process. From (\ref{Anowind}) we 
see that the analog here of the exceptional field theory case would 
be that $\sum_{a}n_{a}\nu_{a}=0$, for in that case a term
with given $n$ and $m$ would not be suppressed as $N\to\infty$.
This, however, can only happen on a set of measure zero 
in the $\nu_{a}$ integration region.
As usual, then, it is the integration over the 
moduli which is responsible for the qualitative difference between the 
string and field theory cases.
The absence of a restriction is clearly related to
the finiteness of the amplitude in the near DLCQ limit, which 
was shown in \cite{Bi} to hold even for processes with no longitudinal 
momentum transfer. 

Our motivation for studying the string theory amplitude was to 
determine whether the presence of objects which can wind around the 
compact direction caused the IMF amplitude to depend on the 
compactification radius.
Even though the preceding analysis was restricted to the case without 
{\em external} winding, it explicitly includes the effects of winding 
modes running in the loop. 
Still, one might worry that the $R$-independence of the amplitude that
was found above could be somehow spoiled in the presence of 
external winding. We thus proceed to examine that case.

Consider then a $2 \to 2$ scattering process with external right- and 
left-moving momenta as in (\ref{kR,L}), now with $m_{a}\neq 0$.
The level-matching condition, $N_{R}-N_{L}=n_{a}m_{a}$, forces the 
vertices to be more complicated than (\ref{vertex}). In fact, 
since the passage to the IMF leaves the winding numbers $m_{a}$ untouched,
as $N \to \infty$ one must consider states which are infinitely 
excited\footnote{This is not merely a 
consequence of the boost; after all, $n_{a}m_{a}=\ap (k_{a L}^{2}-k_{a 
R}^{2})/4$, $N_{R}$ and $N_{L}$ are Lorentz invariant. 
Even in \Fp\ the state of interest 
depends on \eps\ and has infinite oscillator number as $\eps\to 0$. 
More on this later.}.  
The calculation is consequently more intricate than that with 
no external winding. We will analyze here the amplitude obtained
in the simpler setting of the bosonic string, which for this purpose
should be conceptually the same. Coordinate indices are now split into
$M=(0,i,25)$, with $i=1,\ldots,24$, or 
$M=(1,\mu)$, with $\mu=0,2,\ldots,25$.

We must now compute
\bea \label{Adefbos}
A_{\mathrm{cl}}^{(4)} &=&
   \left(\frac{\kappa}{4\pi}\right)^{4}\int\frac{d^{2}w}{|w|^{4}} 
   \int\prod\limits_{a=1}^{3}\frac{d^{2}\rho_{a}}{|\rho_{a}|^{2}} \ I,
   \mbox{\qquad where}\\
I &=&\frac{1}{R}\sum\limits_{n,m=-\infty}^{+\infty}
   \int d^{25}k^{\mu}\ 
   \mbox{Tr}\left[V_{1}(k_{1},\rho_{1})\cdots V_{4}(k_{4},\rho_{4})
   w^{L_{0}}\bar w^{\tilde L_{0}}\right]. \nonumber 
\eea
We choose to satisfy the level-matching condition by using vertices 
with $N_{L}=0$, $N_{R}=n_{a}m_{a}$. The simplest choice (still 
in LC gauge) is 
\be \label{vertexbos}
V_{a}(k_{a},z)= C_{h_{a}}
\left[ \left(\dot{X}_{R}^{a+1}(z) \right)^{h_{a}}
\exp\left(ik_{a R}\cdot X_{R}(z)\right)\right]
\left[
\exp\left(ik_{a L}\cdot X_{L}(\bar z)\right)\right],
\ee
where 
$\dot{X}_{R}^{j}(z)=dX_{R}^{j}(z)/d\sigma_0 =iz\partial_{z}X_{R}^{j}(z)$,
$h_{a}=n_{a}m_{a}$, and $C_{h_{a}}$ is a normalization constant. 
The superscript
$a+1$ above is a coordinate index; the notation indicates
that each vertex is polarized along a different
(noncompact) direction. The
frame of the calculation is chosen to be oriented such that the 
polarization of each vertex is orthogonal to the momenta of all
vertices ($k_{a}^{b+1}=0 \  \forall \ a,b$).

The calculation yields
\bea \label{Abos}
A_{\mathrm{cl}}^{(4)} &=& C
  \int\frac{d^{2}\tau}{(\mathrm{Im}\tau)^{2}}
  \int\prod\limits_{a=1}^{3}\frac{d^{2}\nu_{a}}{\mathrm{Im}\tau} 
  \left( \frac{|f(w)|^{-48}}{(\mathrm{Im}\tau)^{8}|w|^{2}}\right)
  \left( \prod_{a} \calB_{h_{a}}(\tau) \right)\ \calI \\
\calI &=& 
  \prod\limits_{a < b}
  \left[E(\nu_{ab},\tau)^{\ap k_{a R} \cdot k_{b R}/2}
  \bar E(\nu_{ab},\tau)^{\ap k_{a L} 
  \cdot k_{b L}/2} \right]   \nonumber\\
{}&{}&\times \quad \exp\left\{-\frac{\pi \ap}{4\mathrm{Im}\tau}
  \left[\sum\limits_{a}(k_{a R}\nu_{a}-k_{a L}
  \bar\nu_{a})\right]^{2}\right\} \calS   \nonumber\\
\calS &=& \frac{\ap}{R^{2}}\sum\limits_{n,m}
  \exp\left\{-\frac{\pi \ap}{R^{2}\mathrm{Im}\tau}
  (m+n\tau +R\sum\limits_{a}k^{1}_{a R}\nu_{a})
  (m+n\bar\tau 
  +R\sum\limits_{a}k^{1}_{a L}\bar\nu_{a})\right\}. \nonumber
\eea

Here $f(w)=\prod_{j>0}(1-w^{j})$, and all constants have been lumped into
$C$. The factor inside the first large parentheses
is standard for the bosonic string; the second factor results from
the highly excited vertices. $\calB_{h_{a}}(\tau)$ is a 
complicated function of $\tau$ alone (even though one might have
expected it to depend on the insertion points), whose precise form is 
unimportant for the present discussion.

To interpret (\ref{Abos}) it will be important to understand how the 
momenta $k_{a R,L}^{M}$ scale in the limit of interest. In the IMF \Fe, a 
state with longitudinal momentum number $n>0$ and winding number $m$ 
has (right- and left-moving) energy and momentum
\be \label{k}
{k_{R,L}^{0} \choose k_{R,L}^{1}} =
 {{\frac{n}{R} + \frac{\vec{k}_{\bot}^{2} + M^{2}}{2n/R}}
   \choose {\frac{n}{R} \mp \frac{mR}{\ap}}}, 
\ee
where for the bosonic string $M^{2}=(mR/\ap)^{2}+2(N_{R}+N_{L}-2)/\ap$, 
and we have ignored in $k^{0}$ terms of $\calO(\eps^{2})$. 
ln the LFQ frame \Fp\ this corresponds to a state with 
(right- and left-moving) light front energy and momentum 
\be \label{k'} 
{{k'}_{R,L}^{-} \choose {k'}_{R,L}^{+}} =
{{\frac{\vec{k}_{\bot}^{2} + M^{2}}{2n/R'} \mp \frac{mR'}{\ap}}
 \choose {\frac{n}{R'}}}
\ee
(again ignoring $\calO(\eps^{2})$ terms).
Now, for a longitudinally wound string, it is not the momentum ${k'}^{M}$,
but the momentum density ${k'}^{M}/R'$, that should be held fixed as $\eps 
\to 0$. Since in the compactified IMF limit $R'=R/\eps\to\infty$, 
we require $n \propto \eps^{-2} \to \infty$, and 
$\vec{k}_{\bot}\propto \eps^{-1} \to \infty$. The (Lorentz invariant)
level-matching condition $N_{R}-N_{L}=nm$ then implies that at least 
$N_{R}\propto n \to \infty$. Thus, the finite $R$-dependent 
contribution to the mass of the string becomes irrelevant for large 
$N$, and both the mass $M$ and the light front energy 
${k'}^{-}=({k'}_{R}^{-}+{k'}_{L}^{-})/2$ scale as $R'$
as $N \to \infty$. This is appropriate for a string wound around a
circle of radius $R'$.
It had been previously emphasized by Susskind \cite{Su} that 
longitudinally wound objects in  DLCQ ($R=0$) 
decouple as $R' \to \infty$. The present discussion shows that this 
result is independent of $R$.

As a result of the scaling just discussed, the phase of the amplitude
given by (\ref{Abos}) oscillates as $N \to \infty$.
For instance, the piece of the amplitude involving $E$ 
and $\bar E$ can be written as
\be \label{E}
\prod\limits_{a<b}\left|E(\nu_{ab},\tau)\right|^{
  \ap(k_{aR}\cdot k_{bR}+k_{aL}\cdot k_{bL})/2} 
\left(\frac{E(\nu_{ab},\tau)}{\bar{E}(\nu_{ab},\tau)}\right)^{
{\ap(k_{aR}\cdot k_{bR}-k_{aL}\cdot k_{bL})/4}}.
\ee  
The second factor is a pure phase, whose exponent is $\propto 
n_{a}m_{a} \to \infty$.  The amplitude is consequently 
ill-defined in the limit of interest. 
This behavior is a generic property of string amplitudes 
with external winding in this limit; it can be seen to hold for the 
corresponding tree-level amplitude, for example.
The reason is clear: even in the ordinary reference frame \Fp\
the scattering process involves objects of infinite light front 
energies ( $\propto R' \to \infty$).  

Now, the central point for our purpose is that
all $R$-dependence in (\ref{Abos}) disappears as $N\to\infty$. In 
(\ref{E}), 
for example, only the exponent of the first factor depends on $R$. In 
the compactified IMF limit, this exponent is proportional to
\be
  \vec{k}_{a \bot}\cdot \vec{k}_{b \bot}
 -\frac{n_{b}}{n_{a}} \frac{\vec{k}_{a \bot}^{2}+M_{a}^{2}}{2}
 -\frac{n_{a}}{n_{b}} \frac{\vec{k}_{b \bot}^{2}+M_{b}^{2}}{2}
 -\left(\frac{\vec{k}_{a \bot}^{2}+M_{a}^{2}}{2n_{a}/R}\right)
  \left(\frac{\vec{k}_{b \bot}^{2} + M_{b}^{2}}{2n_{b}/R}\right)
 +\frac{m_{a}m_{b}}{\ap^{2}}R^{2}.
\ee 
The last two terms are finite as $N\to\infty$, and therefore become 
irrelevant compared to the rest of the terms, which scale like $N$.
Put differently, Lorentz invariant terms like $k_{a R}\cdot k_{b R}$
can be evaluated in \Fp, where as discussed above they depend on $R'$,
not on $R$. 
Also, the explicit factor of $R^{-2}$ in front of the sums in $\calS$ is,
as in the case with $m_{a}=0$, interpreted as $N^{-2}(N/R)^{2}$: it
is necessary for turning the sums into integrals. 
Though highly formal
because of the ill-defined nature of the amplitude, 
this discussion does make it clear that the amplitude with external 
winding is also consistent with the interpretation of 
$N, R'\to\infty$ as a decompactification limit.

To summarize, then, we have found that in the compactified IMF limit, 
$N \to \infty$ with $R$ fixed, a string one-loop amplitude
becomes independent of $R$, and coincides with that of the
uncompactified theory. 
Unlike the field theory case, this result involves no special 
kinematic restrictions on the scattering process. Moreover,
the features from which the result is infered appear not to be
specific to the one-loop amplitude under consideration. 
One would consequently expect all IMF string amplitudes to
display the same behavior.

\section{M-theory and the Matrix Model} \label{Msec}

We have discovered in Section \ref{fieldsec} that
generic field theory scattering amplitudes are independent of the
longitudinal compactification radius $R$ in the $N \to \infty$ 
limit, but there exist some special processes
(namely, those with no longitudinal momentum transfer between two
subsets of the initial and final particles) for which this is not
true.
In Section \ref{stringsec}, 
string theory amplitudes were found to be independent 
of $R$ in the same limit. 
There are in this case no additional kinematic restrictions 
comparable to the field theory case.
In retrospect, all of this seems quite reasonable. 

What can we learn from this for the case of M-theory? The standard
expectation would be that M-theory should display behavior similar
to that of string theory. The latter is, after all, a special 
ten dimensional limit of the former.
Taken at face value, then, our results indicate that scattering 
amplitudes in M-theory should be independent of $R$ in the IMF.

To see the possible implications of this for the Matrix 
model proposal, let us revisit the scaling arguments of \cite{Se}.
Following standard practice, we will refer in this section
to the compact longitudinal direction in \Fe, of radius $R$,
as the eleventh dimension. 
The relation between quantities in frames
\Fe\ and \Fp\ is given in Fig.~5.

\begin{figure}[ht]  \label{scaling}
 \begin{center} 
  \setlength{\unitlength}{1cm}
 \begin{picture}(15,2.1)
   \put(1.0,0.0){\shortstack[c]{\Fe\\$R$\\$P^{11}=N/R$\\
  $E_{\mathrm{kin}}=R(\vec{P}_{\bot}^{\,2}+M^{2})/2 N$}}
   \put(9.0,0.0){\shortstack[c]{\Fp\\$R'=R/\eps$\\$\Pup=N/R'$\\
   $\Pum=R'(\vec{P}_{\bot}^{\,2}+M^{2})/2 N$}}
   \put(8.8,1.4){\vector(-1,0){3.2}}
   \put(6.6,1.1){$\scriptstyle e^{\omega}=\sqrt{2}/\eps$} 
  \end{picture}
  {\small \caption{Relation between M-theory in the IMF \Fe, and 
     the LFQ frame \Fp. The arrow indicates a 
     longitudinal boost.
     Quantities are given in the {\em same} units in both frames.}}
   \end{center}
\end{figure}

Following \cite{Se}, we notice from Fig.~5 that states with finite
light front energy $\Pum$ in \Fp\ are obtained by holding $R'/N$ fixed.
In \Fe\ such states have kinetic energy 
$E_{\mathrm{kin}}=P^{0}-N/R \propto \eps$.
It is convenient to change in \Fe\ to $\eps$-dependent  
units in which these kinetic  energies are held fixed.
For any quantity $Q$, with mass dimension
$d_{Q}$, we let $Q \to \tilde{Q}= \eps^{-d_{Q}}Q$, where $\tilde{Q}$ 
is the same quantity in the new, changing units.

By definition M-theory with (eleven dimensional) 
Planck length \tlP\
on a spatial circle of radius \tR\ 
is type IIA string theory, with string length and coupling constant
\be
 \tls=  \tR^{-1/2}\tlP^{3/2}=\eps R^{-1/2}\lP^{3/2},\qquad
 \tgs=(\tR / \tlP )^{3/2}=(R/\lP )^{3/2}.
\ee

We now examine M-theory in the three different $\eps \to 0$ limits 
discussed in Section \ref{boostsec}. 
The relation between $N$ and \eps\ in each limit 
is specified there; it depends on whether the objects of interest are 
localized or spread in the longitudinal direction. 
The analysis to follow applies equally well to both cases.

\subsection*{\normalsize Seiberg-Susskind (near DLCQ) limit}

Here we take $\eps \to 0$ holding $N,R'$ fixed.
Then we have M-theory with Planck length $\tlP\propto\eps\to 0$
on a circle with radius
$\tR\propto\eps^{2}\to 0$, i.e., type IIA string theory (in a sector 
with $N$ D0-branes) with coupling constant 
$\tgs \propto \eps^{3/2} \to 0$ and string
length $\tls \propto \eps^{1/2} \to 0$. For any transverse
distance $r_{\bot}$  we have $\tilde{r}_{\bot}/\tls\propto 
\eps^{1/2} \to 0$.  As explained in \cite{Se}, the physics of the
theory in this limit is correctly described by the Matrix model. 
However, as pointed out in \cite{Ba,HP}, and discussed in 
sections 1 and 2,
it is not clear that the full eleven dimensional theory
could be recovered after having taken this limit, even if
$N, R' \to \infty$ afterwards.

\subsection*{\normalsize BFSS (uncompactified IMF) limit}
 
Now $\eps\to 0$ with $N \to \infty$, 
$R\propto \eps^{-\alpha}\to\infty$, where $0<\alpha<1$ (see Section 
\ref{boostsec}).
Then $R'\propto\eps^{-1-\alpha}\to \infty$. 
This is M-theory with Planck length $\tlP\propto\eps\to 0$ on a 
circle of radius $\tR\propto\eps^{1-\alpha}\to 0$, i.e., type IIA 
string theory (in a sector with $N \to \infty$ D0-branes) with 
coupling constant $\tgs \propto \eps^{-3\alpha /2}\to\infty$ and 
string length $\tls \propto \eps^{1+\alpha /2}\to 0$. The vanishing 
of \tls\ causes the string oscillators to decouple, just like in the 
previous limit. Also, we are certain that we are examining here 
the uncompactified M-theory. There is a price to pay, however, 
for this certainty: the string theory is now strongly coupled, and 
as a result the Matrix model cannot be justified. In 
addition, for any transverse distance $r_{\bot}$ 
we now have $\tilde{r}_{\bot}/\tls\propto\eps^{-\alpha/2}\to\infty$,
so we are in the supergravity ($r_{\bot}\gg \ls$),
{\em not} the Matrix theory  ($r_{\bot}\ll \ls$) regime \cite{DKPS}. 
All of this
is no surprise, of course: uncompactified M-theory is by definition 
the strong-coupling dual of type IIA string theory. This is why the 
BFSS proposal was simply a conjecture.
 
\subsection*{\normalsize Compactified IMF limit}

Here $\eps\to 0$ with $N \to\infty$, $R$ fixed. Then
$R'\propto\eps^{-1}\to\infty$. This is M-theory
with Planck length $\tlP\propto\eps\to 0$ on a circle of radius 
$\tR\propto\eps\to 0$, i.e., type IIA string theory 
(in a sector with $N \to \infty$ D0-branes) with {\em finite}
coupling constant $\tgs=(R/\lP)^{3/2}$ and vanishing string length
$\tls\propto\eps\to 0$. The string oscillators still decouple. 
The string theory can now be weakly or strongly coupled, according to 
whether $R\ll\lP$ or $R\gg\lP$. Similarly, for any transverse 
distance $r_{\bot}$ we have $\tilde{r}_{\bot}/\tls 
=(r_{\bot}/\lP)(R/\lP)^{1/2}$, so the size of $R/\lP$ also determines 
whether we are in the Matrix theory  ($r_{\bot}\ll \ls$) or 
supergravity ($r_{\bot}\gg \ls$) regimes.

\bigskip

Under the assumption that all limits commute, both the 
Seiberg-Susskind limit followed by $N,R' \to\infty$,
and the BFSS limit, are continuously connected to the 
cases with a finite value of $R$. Even if this is not true, 
the compactified IMF limit certainly includes the cases with arbitrarily
small and large $R$.

Our work suggests that M-theory scattering amplitudes 
are independent of $R$ in the large $N$ limit. If correct, 
this would explain how Matrix theory, obtained as a description of 
ten dimensional processes, is able to encapsulate information about 
an eleven dimensional theory. As emphasized in the introduction, this is 
unquestionably a mysterious issue in the Matrix program, 
particularly because in string theoretic 
language it involves an {\it a priori}\/ 
unwarranted extrapolation from weak 
to strong coupling. The Matrix model can only be 
satisfactorily justified for small $R$ \cite{Se}, and yet one expects to 
extract from it a description of eleven dimensional physics.
$R$-independence of amplitudes in the IMF might be the key to
understand why this is not contradictory. The point is simply that
it is not necessary to take $R \to \infty$ (as in \cite{BFSS}) to obtain the
uncompactified theory. If M-theory in the IMF is indeed independent 
of $R$, then one can, in that frame, simultaneously interpret it as 
weakly coupled type IIA string theory (with $N \to \infty$ D0-branes), 
or as uncompactified eleven dimensional M-theory!

To be fair, we should point out that a skeptic could still adopt just the
opposite view: the fact that, for any $R$, as $N \to \infty$ one 
obtains the uncompactified theory, could be taken to  
invalidate Seiberg's interpretation of M-theory on a vanishingly small
circle in the IMF as weakly coupled type IIA string theory. If this
is the case, the remarkable success of Matrix theory 
still remains to be explained.

Balasubramanian, Gopakumar and Larsen \cite{BGL} have provided evidence
for the possibility of recovering the eleven dimensional theory
from the near DLCQ ($R\to 0$) limit followed by $N,R'\to\infty$.
Our results likewise substantiate the interpretation of 
$N,R'\to\infty$ as a decompactification limit (for any value of
$R$). This appears to discard then the 
suggestion made in \cite{Ba} that perhaps the discrepancies reported 
in \cite{DR} could be an indication that the near DLCQ theory does not 
converge to the uncompactified theory for large $N$.

The arguments of \cite{BGL} additionally support the conjecture 
that M-theoretical physics is independent of $R$ in the large $N$ limit. 
The authors of \cite{BGL} have in fact also realized this, and cite as direct
evidence for the conjecture the agreement between the effective 
action of a D0-brane probe-target system computed in eleven dimensional
light front supergravity and in string theory at disk level. 
They also remark that, if the 
conjecture were correct, loop corrections to this process
in string theory (due to handles and holes) should have specific
Matrix model counterparts.

On the other hand, our work also shows that the question of 
$R$-independence might be subtle. In field theory, amplitudes 
for scattering processes with no longitudinal momentum transfer 
between two subsets of the initial and final particles depend on 
$R$ even in the large $N$ limit. 
The calculations of Matrix model amplitudes have been almost
exclusively restricted to precisely this case\footnote{Incidentally,
Matrix theory amplitudes in the literature are expressed in terms of
a potential, the Fourier transform of the momentum space amplitude.
For processes with no longitudinal momentum transfer the Fourier transform
introduces an additional overall factor of $1/R$ in \Fe, or of
$1/R'$ in \Fp.}, due to our ignorance of the
necessary bound state wavefunctions. It is conceivable
that this restriction could complicate the comparison with supergravity
one-loop amplitudes,
which might behave like the exceptional field theory case, retaining
some $R$-dependence even for $N\to\infty$. 

Of course, as long as we lack a definitive fundamental formulation of
M-theory, analogies to string and field
theory are not guaranteed to be perfect. 
The best attempt at a 
definition of M-theory to date is the Matrix model itself, the status
of which these analogies are meant to shed light on.
The description
of states and interactions in Matrix theory is so different
from that of conventional field and string theory (especially
with respect to locality, the nature of the fundamental objects,
and the short-distance structure of spacetime) that
any evidence based on analogies to these other cases can at best
be regarded as indirect.

\section{Acknowledgements}

It is a pleasure to thank Curtis Callan, Diego C\' ordoba, Sangmin Lee,
Sanjaye Ramgoolam, Anastasia Ruzmaikina, \O yvind Tafjord and 
L\' arus Thorlacius for useful discussions. 
I am also grateful to Curtis Callan for critical review of 
the manuscript.
This work 
was partially supported by DOE grant DE-FG02-91ER40671,
and by
the National Science and Technology Council of Mexico (CONACYT).

\end{document}